\documentclass[twocolumn,
aps,nofootinbib,showpacs,showkeys,preprint
tightenlines,preprintnumbers,] {revtex4}

\usepackage{epsf,epsfig,subfigure,graphicx,amsmath,amssymb}
\usepackage{color}
\newcommand{\dis}[1]{\begin{equation}\begin{split}#1\end{split}\end{equation}}

\newcommand{\etal}{{\it et al.}}

\newcommand{\gev}{\,\textrm{GeV}}

\newcommand{\thet}{{\theta\,}}

\newcommand{\Z}{{\bf Z}}

\newcommand{\ie}{{\it i.e.~}}

\newcommand{\Vde}{{$10^{-47}\,\gev^{\,4}$}}

\newcommand{\UDE}{{U(1)$_{\rm de}$}}
\newcommand{\UPQ}{{U(1)$_{\rm PQ}$}}

\newcommand{\MG}{{M_{\rm GUT}}}

\newcommand{\fde}{{f_{\rm DE}}}

\begin{document}
\draft

\title{\Large\bf The inflation point in U(1)$_{\rm de}$ hilltop potential assisted by chaoton,   BICEP2 data, and trans-Planckian decay constant}

\author{  Jihn E.  Kim }
\affiliation{
 Department of Physics, Kyung Hee University, 26 Gyungheedaero, Dongdaemun-Gu, Seoul 130-701, Republic of Korea
}

\begin{abstract}
The recent BICEP2 report on the CMB B-mode polarization hints an early Universe energy density at the GUT scale. We add a new `chaoton' term to our recently proposed hilltop potential to have a large tensor mode fluctuation. The chaoton field slides down from the hilltop when the inflaton field value is small so that an enough $e$-folding is possible.
We also comment how the trans-Planckian decay constant is obtained from some discrete symmetries of ultra-violet completed models.

\end{abstract}


\maketitle

\section{Introduction}

The recent report of the tensor modes on a large CMB B-mode polarization by the BICEP2 group
\cite{BICEP2I} has attracted a great deal of attention. The reported tensor-to-scalar ratio is $r=0.2^{+0.07}_{-0.05}$ (after dust reduction to  $r=0.16^{+0.06}_{-0.05}$). But, the previously reported Planck data presented an upper bound on $r<0.11$ \cite{Planck13} which is about $2\sigma$ away from the BICEP2 report. At present, therefore, we need to wait a final confirmation on the BICEP2 report. However, this large value of $r$ is so profound if true, here we investigate a possible outcome from our recently published {\it hilltop inflation} model \cite{KimNilles14, KimJK14}.

\begin{figure}[!t]
\begin{center}
\includegraphics[width=0.75\linewidth]{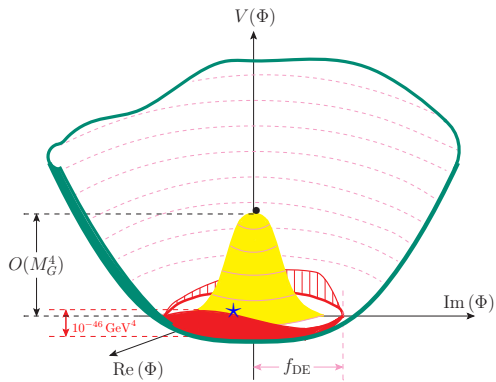}
\end{center}
\caption{The dark energy potential. The blue star marks a typical field value of the phase field of $\Phi$.  } \label{fig:DEpot}
\end{figure}

A large $r$ seems against hilltop inflation scenario rolling down from the origin \cite{Buch14}. However, a hilltop potential is quite generic from the top-down approach \cite{Busso13}.
In Ref. \cite{KimNilles14}, the hilltop inflation was suggested on the way to understand a very tiny dark energy (DE) scale \Vde~  \cite{Perlmutter98,Riess98}, by closing the shift symmetry $a_{\rm de}\to a_{\rm de}+{\rm constant}$ of the DE Goldstone boson direction. The field $a_{\rm de}$ is a pseudo-Goldstone boson because any global symmetry must be broken at some level \cite{KimPLB13, KimNilles14}. For $a_{\rm de}$ to generate the DE scale, theory must allow the leading contribution to DE density at the level of $10^{-46\,}\gev^4$. A top-down approach such as string theory introduces the defining scale ($M_P\simeq 2.44\times 10^{18}\,\gev$ or string scale), and the next possible scale is the grand unification (GUT) scale $\MG $. If $a_{\rm de}$ is a pseudo-Goldstone boson with its decay constant at a Planckian (or trans-Planckian) value, its potential can be parametrically expressed as a power series of $\MG/M_P$. However, if  $a_{\rm de}$ couples to the QCD anomaly, then it is the QCD axion.\footnote{We can neglect the coupling to the SU(2)$_{\rm weak}$ anomaly, whose effect is negligible compared to the potential energy term we consider as powers of  $\MG/M_P$.} Since the QCD axion cannot be $a_{\rm de}$, we must introduce two spontaneously broken global symmetries, one \UPQ~ and the other \UDE, where \UDE~ is chosen not to carry the QCD anomaly. If the leading term of $a_{\rm de}$ is chosen at the $10^{-46\,}\gev^4$ level, its potential looks like Fig. \ref{fig:DEpot}, where this tiny energy scale is shown as the red band (exaggerated in the figure), and the decay constant of $a_{\rm de}$, $\fde$, can be larger than the Planck mass $M_P\simeq 2.44\times 10^{18}\,\gev$.
The decay constant $\fde$ is required to be trans-Planckian so that $a_{\rm de}$ has survived until now \cite{Carroll98}. One inevitable aspect of this study is that it is necessary to consider \UDE~ (and hence the QCD axion \cite{InvAxionRev10}) together with the U(1)$_{\rm de}$ symmetry.

The field $a_{\rm de}$ is a pseudoscalar field, \ie the phase of some complex scalar $\Phi$. In the top-down approach, the height of the potential at the origin is expected to be of order $\MG^4$ as shown in Fig. \ref{fig:DEpot}. Since $a_{\rm de}$ is the phase of $\Phi$, the potential along the $a_{\rm de}$ direction is flat if we do not consider the explicit breaking terms of order $10^{-46\,}\gev^4$. Of course, at the intermediate scale or at the electroweak scale, there are additional \UDE~ breaking terms, but their effect is just changing $\fde$  by a tiny amount, $\fde\to\sqrt{\fde^2+M^2_{\rm int}}$. In this top-down approach, we must consider the potential shown in Fig.  \ref{fig:DEpot}, and the very early Universe might have started at the black bullet point of Fig.  \ref{fig:DEpot} due to high temperature effects \cite{Guth,NewInfl}. This leads to the hilltop inflation. Our `hilltop inflaton' is a scalar field.

The `natural inflation' \cite{NaturalInfl,Peloso04} is also using a potential of a pseudo-Goldstone boson, but it is not a hilltop inflation because at the origin of Im\,$(\Phi)$ the potential is a local minimum and the `natural inflaton' is a pseudoscalar field.     Nevertheless, the possibility of large $r$ from the BICEP2 data may rule out the hilltop inflation, even though a $(3-4)~\sigma$ allowance may be acceptable. On the other hand, if the height of hilltop is much lower than the GUT scale energy density, the inflation history may not be affected by the hilltop potential. A more attractive possibility will be that the inflaton may not be a vanilla type single field but involves more than one field.

In the Einstein equation $G_{\mu\nu}=T_{\mu\nu}$, the Einstein tensor responds to the energy-momentum tensor and the GUT energy density can be considered small enough to use the Einstein equation for the evolution of the Universe. If there exists a trans-Planckian vacuum expectation value (VEV) or decay constant, one should check a possible generation of Planck scale $T_{\mu\nu}$ in which case a proper discussion of the Universe evolution by the Einstein equation is impossible. But, if the energy scale during inflation is small (\ie  $(10^{16}\,\gev)^4$) compared to the Planck energy density $M_P^4$, the trans-Planckian field values (\ie the DE decay constant $\fde>M_P$) are  allowed during inflation \cite{Lyth97}.

One possible trans-Planckian decay constant is some combination of axion decay constants \cite{Peloso04,Nflation08} where the potential energy never exceeds $M_P^4$ due to the shift symmetries of axions. The form of the potential of Fig. \ref{fig:DEpot} is also appropriate for inflation if we let $|\Phi|<\fde$.
Usually, the cutoff scale of Planck mass allows higher dimensional operators $\phi^n/M_P^{n-4}$ for field value of $\phi$ less than the cutoff scale. With the trans-Planckian $\fde$, it corresponds to $\frac{\lambda\phi^n}{M_P^{n-4}}(\textrm{the vacuum energy at }\phi=0)< M_P^4$, or the trans-Planckian decay constant satisfies, $\fde <M_P /\lambda^{1/n}$. This corresponds to allowing only smaller and smaller couplings for higher order terms of $\phi$ such as $\cos\phi$ \cite{NaturalInfl,Peloso04}. We will also point out that even without shift symmetries an appropriate choice of discrete quantum numbers of the inflaton and GUT scale fields can be adequate to describe a trans-Planckian VEV of the inflaton.

\section{Spontaneously broken U(1) hilltop inflation}
\begin{figure}[!t]
\begin{center}
\includegraphics[width=0.85\linewidth]{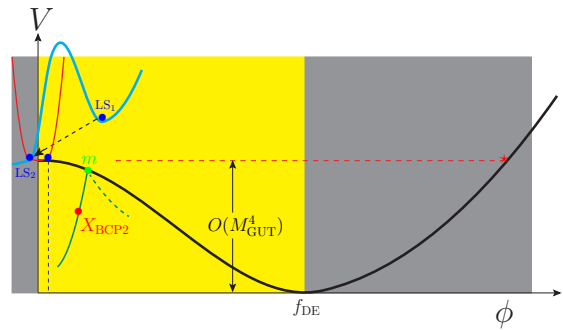}
\end{center}
\caption{The \UDE\,-hilltop inflation. The cyan curve is the potential showing tunneling to the blue bullet. The blue bullets in the gray and yellow are the equivalent points. The temperature dependent potential before spontaneous symmetry breaking of \UDE\, is shown as the red curve. The green curve direction from $m$, orthogonal to that of $\phi$, is the chaoton direction.}\label{fig:Hilltop}
\end{figure}

Let us introduce dimensionless energy variables in units of $M_P\simeq 2.44\times 10^{18}\,\gev$ and a dimensionless time $t$ in units of $M_P^{-1}$. A GUT scale reported in Ref. \cite{BICEP2I} is $(2\times 10^{16}\,\gev)^4$ which is about $ 10^{-8} $.
Models from (heterotic-)string compactifications leading to the unification of gauge couplings at the GUT scale \cite{Kim:2006hw, Lebedev:2006kn, Kim:2006zw, Lebedev:2007hv, Lebedev:2008un, Nilles:2008gq} do not necessarily imply renormalizable couplings in the effective potential $V$ below the Planck scale $M_P$. There are two well-known hilltop forms for the potential, which are very flat near the top.

The first example is the quartic potential with an extremely small $\lambda$. With the symmetry $\phi\to -\phi$, it can be written with two parameters, $\lambda$, and $\fde$, with three conditions, $ V'(0)= 0, V'(\fde)=0 $,  and $V(\fde)=0$,
\dis{
V =\frac{\lambda M_P^4}{4!}\, (\phi^2 -\fde^2)^2 \equiv \frac{\lambda  }{4!}\, (\phi^2 -\fde^2)^2  \label{eq:Ftermpotential} 
}
where $\lambda$ is the quartic coupling constant and $\phi$ is the radial field of  Fig. \ref{fig:DEpot}.

The second example is the non-supersymmetric Coleman-Weinberg (CW) type potential \cite{CW82,WeinbergCW76}, originally considered in the new inflation scenario \cite{NewInfl},
\dis{
\textrm{CW}\left\{\begin{array}{l}
V=B\left(\phi^4\,\ln \frac{\phi^2}{M_f^2} +\frac12  e^{-1} M_f^4 \right) , \\[0.6em]
V'=4B\phi^3\left(\ln \frac{\phi^2}{M_f^2}+\frac12\right) ,\\[0.6em]
 V^{\prime\prime}=12B\phi^2\left(\ln \frac{\phi^2}{M_f^2}+\frac76\right).\end{array}
\right.\label{eq:CWpotential}
}
where $M_f$ is a mass parameter chosen to absorb all $\phi^4$ coupling in $V(\phi)$,   and
\dis{
B=\frac{3}{64\pi^2 \phi^4}{\rm Tr}\,\mu_\phi^4=\frac{3}{64\pi^2 \langle \phi\rangle^4}\sum_{v}\mu_v^4
}
where for simplicity we did not include the fermion and scalar couplings and the sum running over all massive vector bosons at the GUT scale. With the  CW potential, it is known that the Higgs mass is O($\alpha$) times smaller \cite{WeinbergCW76} than the VEV of the Higgs field. In the \UDE\,case, the VEV or $\fde$ is required to be trans-Planckian and a GUT scale scalar mass perfectly fits with a trans-Planckian DE decay constant. If the BICEP2 data is explainable with the CW potential, it is a very attractive one relating the scales of $\fde$ and $M_{\rm GUT}$.
There are more examples of inflatons, mostly with large field values for inflation.

A year ago the small field inflation was looked plausible with the Planck data \cite{Planck13}, possibly disfavoring a large field value, but the situation has changed after the report by the BICEP2 group.
In each case, Eq. (\ref{eq:Ftermpotential}) or Eq. (\ref{eq:CWpotential}), the potential is schematically drawn in Fig. \ref{fig:Hilltop}. But, there is a problem with the hilltop potential with a large $r$ if inflation starts from the origin.
This is because with the BICEP2 value of $r$, $1-\frac38 r\simeq 0.925$. With Eqs.  (\ref{eq:Ftermpotential}) and (\ref{eq:CWpotential}), we have a very small $\eta$, and the relation $n_s=1-\frac38 r+2\eta$ cannot be raised to $\sim 0.96$. This is even before calculating the $e$-folding number in a specific inflation model.
Hence, central values of both $r$ and $n_s$ cannot be explained at the same time with the \UDE\,hilltop model. Thus, we must include additional terms if  \UDE\,hilltop height is
contributing to the slow roll inflation in a nontrivial way.

\section{Addition of chaoton field $X$}
\begin{figure}[!t]
\begin{center}
\includegraphics[width=0.75\linewidth]{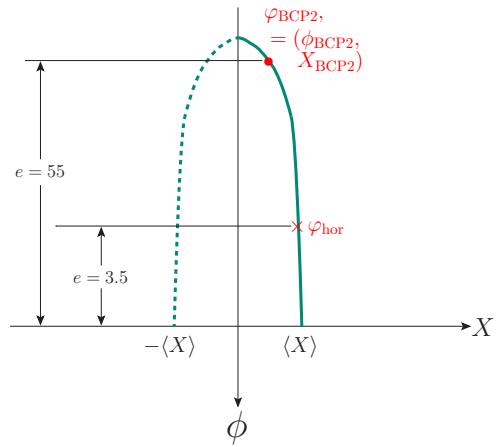}
\end{center}
\caption{The inflation path field $\varphi$ where the direction $\theta$ of Eq. (\ref{eq:PathofInfl}) defines $\varphi$ in the space of two fields, $\phi$ and $X$.   } \label{fig:Path}
\end{figure}

Let us include an additional inflaton field $X$ to locate at bull's eye of the BICEP2 data. This is to mimic the chaotic inflation  \cite{Linde83}, and hence we call $X$  {\it chaoton}. The green curve in Fig. \ref{fig:Hilltop} is the second rolling direction along $X$, and let us consider the following potential
\dis{
V=\lambda_\phi(\phi^2-\fde^2)^2  +\lambda_X\left[X^2-a(\phi^2-m^2)\right]^2 ,\label{eq:Water}
}
where $\lambda_\phi,\lambda_X, a,\fde,m>0 $. When the inflaton $\phi$
moves from 0 to $m$, $X$ stays at 0.  Immediately after $\phi$ passes  $m$, the $X$ moves to nonzero values, either to + or -- direction, as shown in Fig. \ref{fig:Hilltop}.
It is like the hybrid inflation, and we make the slope along $X$ large as soon as $\phi$ passes $m$ so that the chaotic inflation is easily mimicked.
The minimum is now shifted from $\phi=\fde$ to
\dis{
&\langle\phi\rangle =\fde,\\
&\langle X\rangle =\pm \sqrt{a(\fde^2-m^2 )}\,.
}
Somewhere in the second roll, we expect the BICEP2 measure point is located, which we marked as $(\phi,X)=(\phi_{\rm BCP2},X_{\rm BCP2})$. The top view of inflation path is shown in Fig. \ref{fig:Path}.

After passing $\phi=m$, the role of inflaton is changing from $\phi$ to $\varphi$, and the inflation path is a curve in the $X-\phi$ plane
\dis{
&\varphi=\cos\thet \phi-\sin\thet X,\\
&\varphi_\perp=\sin\thet \phi+\cos\thet X,\\[0.5em]
&\phi=\cos\thet \varphi+\sin\thet \varphi_\perp,\\
&X=-\sin\thet \varphi+\cos\thet \varphi_\perp,\label{eq:PathofInfl}
}
where $\theta$ is a function of $t$.
The first and second derivatives are
\dis{
& V_{,\phi}  = 4\lambda_\phi\phi(\phi^2-\fde^2)-4a\lambda_X \phi \left[X^2-a(\phi^2-m^2)\right],\\
& V_{,X} = 4 \lambda_X   \left[X^3-a(\phi^2-m^2)X\right],\\
& V_{,\varphi}=-4\lambda_\phi c_\theta \phi(\fde^2-\phi^2)\\
&\quad\quad~ -4\lambda_X(s_\theta X+a c_\theta\phi) \left[X^2-a(\phi^2-m^2)\right],
}
\dis{
& V_{,\phi \phi}  = 4\lambda_\phi(3\phi^2-\fde^2)-4a\lambda_X \left[X^2-a(3\phi^2-m^2)\right],\\
& V_{,X \phi}  =  -8a\lambda_X  X\phi ,\\
& V_{,X X}  = 4\lambda_X\left[3X^2- a(\phi^2-m^2)\right],\\
& V_{,\varphi \varphi} =4\lambda_\phi c_\theta^2 (3\phi^2-\fde^2)-8\lambda_Xa s_\theta c_\theta \phi X\\
&\quad\quad\quad +4 \lambda_X s_\theta^2 \left[3X^2-a(\phi^2-m^2)\right]
,
}
where
\dis{
\tan\thet= \frac{X}{\phi}.
}
Along the inflaton direction $\varphi$, the slope is
\dis{
&V_{,\varphi}=V_{,\phi}\cos\thet +V_{,X} \sin\thet.
}
If falling along the green direction is immediate, we can take $\theta$ is close to $\frac{\pi}{2}$ as soon as $\phi$ passes $m$.
Roughly speaking, the path may look like Fig. \ref{fig:Path} such that the chaoton immediately settles to its minimum and the next rolling is mostly via $\phi$. The green path in Fig. \ref{fig:Path} is the inflation path $\varphi$ after  $\phi$ passes $m$. At $\varphi_{\rm BCP2}$, $\theta$ is large but it quickly becomes zero and a slow roll continues for a long time until $\varphi$ reaches  $\varphi_{\rm hor}$. After  $\varphi$ reaches  $\varphi_{\rm hor}$, it will oscillate quickly around the minimum  $\varphi=\fde$, or another waterfall field takes over to end the inflationary epoch.

\section{Sufficient inflation}

The needed $e$-folding number is of order $50\sim 60$. We can take $\varphi$ as the point $(\phi_{\rm BCP2},X_{\rm BCP2})$.  In the slow roll, the $e$-fold number $N$ is given by
\dis{
N(\varphi)\simeq
\int_{\varphi_{\rm end}}^{\varphi}\frac{V(\varphi)}{V (\varphi)_{,\varphi}}d\varphi
\,,\label{eq:efold}
}
along the path $d\varphi=d(\phi\cos\theta+X\sin\theta)$ where $V_{,\varphi}=V_{,\phi}\cos\theta+V_{,X}\sin\theta$. Since $V_{,\phi}$ and $V_{,X}$ are sufficiently flat, we can obtain enough $e$-folding number. Note that $N(\varphi_{\rm hor})$ is of order $3\sim 4$. A possible difficulty for large $N(\varphi_{\rm BCP2})$ from Eq. (\ref{eq:efold})
arises from the fact that the numerator of the integrand is of order GUT scale energy density fixed by BICEP2 while the denominator is in general a function of field value $\varphi$. We can solve this problem numerically by solving \cite{JeongKim14}
\dis{
\ddot\varphi+3H(t)\dot\varphi+\frac{dV}{d\varphi}=0.\label{eq:Evolution}
}

Even before solving Eq. (\ref{eq:Evolution}), we can check possible conditions for a large tensor to scalar ratio $r$  and an appropriate tilt $n_s$, by checking the first and second derivatives of $V$: $V_{,\varphi}$ and $V_{,\varphi\varphi}$,
\begin{widetext}
\dis{
r &=16\epsilon=8\left(\frac{V_{,\varphi}}{V}\right)^2=2 \left(8\, \frac{ [\lambda_XX s_\theta-a\lambda_X\phi c_\theta][X^2-a(\phi^2-m^2)] -  \lambda_\phi c_\theta\phi(\fde^2-\phi^2)  }{\lambda_X [X^2-a(\phi^2-m^2)]^2+
 \lambda_\phi(\fde^2-\phi^2)^2 }\right)^2,
}
\dis{
2\eta= 8\frac{\lambda_\phi c_\theta^2 (3\phi^2-\fde^2)+ \lambda_X s_\theta^2 \left[3X^2-a(\phi^2-m^2)\right]
-2\lambda_Xa s_\theta c_\theta \phi X}{\lambda_\phi(\fde^2-\phi^2)^2  +\lambda_X\left[X^2-a(\phi^2-m^2)\right]^2 },
}
\end{widetext}
where $c_\theta=\cos\theta$ and $s_\theta=\sin\theta$. In the limit $\frac{\lambda_\phi}{\lambda_X} ,a \to 0$, we obtain $r\to 128 s^2_\theta/\langle X\rangle^2$ and $2\eta\to 24 s_\theta^2$. This leads to $|\theta|\simeq 2.2^0$ and $X=0.966$ to have $r=0.2$ and $n_s=0.96$. In this case, the slow-roll gives the integrand of Eq. (\ref{eq:efold}) as $\sim 26$ and we need the slow-rolling continues until $(\varphi_{\rm end}-\varphi_{\rm BCP2})\sim 2$ to have $N\sim (50-60)$. This rough estimate, for the limit $\frac{\lambda_\phi}{\lambda_X} ,a \to 0$, is just to show the existence of possible solutions.
This limit is effectively discounting the \UDE\,hilltop potential compared to the chaoton potential.

For a small $\theta$ region,
\dis{
 &\sqrt{\frac{r}{2}}\simeq  \left|8\phi\,\frac{ \frac{  (\fde^2-\phi^2)}{[X^2-a(\phi^2-m^2)]^2}+\frac{  a(\lambda_X/\lambda_\phi ) }{X^2-a(\phi^2-m^2)}  }{(\lambda_X/\lambda_\phi )+
\frac{(\fde^2-\phi^2)^2}{[X^2-a(\phi^2-m^2)]^2}}\right|,
}
\dis{
2&\eta\simeq  \frac{8(3\phi^2-\fde^2) }{ (\phi^2-\fde^2)^2  +(\lambda_X/\lambda_\phi)\left[X^2-a(\phi^2-m^2)\right]^2 }.
}
For $2\eta$ to be positive in the downhill region, we require $\phi$ to be sufficiently large $\frac{1}{\sqrt3}<\frac{\phi}{\fde}<1$. If the \UDE\,hilltop potential is more significant than the chaoton, the region for the $\lambda_X/\lambda_\phi\to 0$ limit can be considered, in which case we obtain
\dis{
 &\sqrt{\frac{r}{2}}\simeq  \left|\frac{8x}{(1-x^2) }\right|,
~2 \eta\simeq  \frac{8\,(3x^2-1) }{ (1-x^2)^2  }.\label{eq:largePhiCoupl}
}
where $x=\phi/\fde$. For a nonzero $r$, we need $\fde>\sqrt{48}$. But, then Eq. (\ref{eq:largePhiCoupl}) gives a too large $r$ for $\phi>\fde/\sqrt3$. Therefore, a reasonable value of $\theta$ is needed.

\begin{figure}[!t]
\begin{center}
\includegraphics[width=0.95\linewidth]{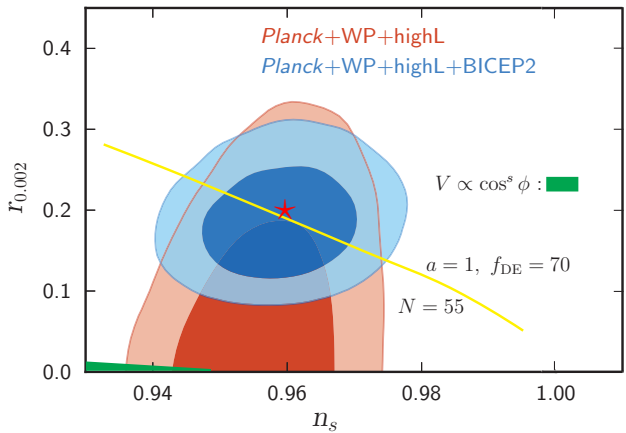}
\end{center}
\caption{The \UDE\,-hilltop estimation of $r$ and $n_s$ over the background-blue tone of
BICEP2 plus Planck data and the background-brown tone of Planck-only data. The red star is the BICEP2 point, $r=0.2$ and $n_s=0.96$. The green region of $n_s\lesssim 0.95$ is for a kind of single field hilltop potential $V\propto \cos^s \phi$ for any $s$. Our model prediction for $N=55$ (the yellow curve) is illustrated with parameters $a=1$ and $\fde=70$ of Eq. (\ref{eq:angle45}).}\label{fig:OnBICEP2}
\end{figure}

For $\phi$ to contribute also significantly in the inflation, we can take  comparable $\lambda_\phi$ and $\lambda_X$, and also a nonnegligible $\theta$. To check this region, let us study $\lambda_X=\lambda_\phi$ and $\theta=\pm\frac{\pi}{4}$. Then,
we have
\dis{
\sqrt{r}\,&= \frac{\left|(X-a\phi)[X^2-a(\phi^2-m^2)]-\phi(\fde^2-\phi^2)\right| }{(\fde^2-\phi^2)^2+ [X^2-a(\phi^2-m^2)]^2 } \\
2\eta &=4\sqrt{r}\,\frac{(3\phi^2-\fde^2) + [3X^2-a(\phi^2-m^2)] \mp 2\phi X }{\left|(X-a\phi)[X^2-a(\phi^2-m^2)]-\phi(\fde^2-\phi^2)\right| }.
}
If $\theta$ turns to $\pm\frac{\pi}{4}$ (for $\pm \phi$ direction) in a short fall of $X$ while $\phi$ has a trans-Planckian shift, \ie $|X/\phi|\ll 1$, we have
\dis{
n_s\simeq 1-\frac38 r+4\sqrt{r}\,\frac{(3-a)\phi^2-\fde^2}{(1+a^2)\phi^3-\fde^2\phi}.\label{eq:angle45}
}
For $\varphi_{\rm BCP2}=1/\sqrt2, r=0.2$ and $n_s=0.96$, we need $\fde(1+a)\simeq 73$, which are the conditions before calculating the $e$-fold number. So, it is possible to satisfy the BICEP2 point. But, there is a problem in obtaining a large $e$-folding. That is our reason that a large $e$-folding is mainly obtained by shifting from a non-negligible  $\theta$ to the $\theta\simeq 0$ direction as shown in Fig. \ref{fig:Path}.
To illustrate a possibility for a particular choice of parameters with the form Eq. (\ref{eq:angle45}), let us choose $a=1$ so that $n_s$ vs. $r$ function takes a simple form $n_s\simeq 1-\frac38 r+(4\sqrt{r}/\phi)$. If we assume most of 50--60 $e$-folding is obtained by the $\phi$ path immediately after the detour along $X$, we have the $e$-folding region as shown in Fig. \ref{fig:OnBICEP2} for $\fde=70$.  But, it is not accurate in the sense that we chose a specific angle and assumed the final path along $\phi$ giving most of $e$-folding. A reliable numerical study is necessary \cite{JeongKim14}.

\section{Trans-Planckian scale from discrete symmetries}

With a single field inflation with a large tensor mode $r$, there is the well-known Lyth bound $ \phi\gtrsim 15\,M_P$ \cite{Lyth97}. Since BICEP2 report indicates a non-negligible $r$, we can use Lyth's trans-Planckian VEV condition to constrain possible theories \cite{Lyth14}. Even if a chaotic inflation with $\phi^2$ can describe a large $r$ with a trans-Planckian VEV, it is required in this case to explain why all higher order terms are neglected. For example, one may consider
\dis{
V_{\rm Lyth}=\frac12 m^2\phi^2 +\sum\frac{\lambda_d}{M^{d-4}}\,\phi^d,
}
from which a slow-roll parameter is calculated as
\dis{
\eta(N)=\frac{m^2}{3H_I^2} +\frac{\sum d(d-1)\lambda_d\phi^d/M^{d-4}}{3H_I^2\phi^2}.
\label{eq:SlowRoll}
}
Barring the accidental cancellation, the condition for a trans-Planckian VEV of $\phi$ is $d(d-1)\lambda_d<6\times 10^{-9}$ \cite{Lyth14}. This is a slow-roll inflation condition. Pictorially, we reinterpret this in Fig. \ref{fig:Hilltop}. Even without the condition for a slow-roll, the Einstein equation, $G_{\mu\nu}=T_{\mu\nu}$, must be applicable in the evolution of the Universe. It amounts to requiring the vacuum energy $V$ in Fig. \ref{fig:Hilltop} must be sufficiently small. The natural inflation has $V$ as the cosine function such that in the gray region of Fig. \ref{fig:Hilltop} it is not going up above $O(M_G^4)$, \ie $V$ is bounded by the red dash-line, and consideration of a trans-Planckian VEV of $\phi$ does not lead to $V$ larger than  $O(M_G^4)$ \cite{Peloso04,Nflation08}.
For example, in string theory one allows all non-renormalizable terms below the string scale $m_s\lesssim M_P$. The question is, ``Why do we neglect a term such as $\phi^{104}/m_s^{100\,}?$ Its coefficient must be smaller than $10^{-127}$ not to disrupt the quadratic term for the dominant contribution to the inflation." This argument applies to all possible terms from string theory. On top of this, one can add here the slow-roll condition (\ref{eq:SlowRoll}), which gives conditions on the coupling constants. But condition (\ref{eq:SlowRoll}) is not that strong compared to those forbidding all possible non-renormalizable terms.

In the hilltop inflation, the inflation region is in the yellow part of Fig. \ref{fig:Hilltop}. In this region, theory is well behaved if $V$ is bounded by $O(M_G^4)$. Here, we add another way to realize the trans-Planckian decay constant along this line of argument. In the yellow hilltop region, the ratio $\phi/M_P$ has a fixed value even though it can be $O(10)$.

 Suppose string theory allows a $\Z_N$ or $\Z_{nR}$ symmetry \cite{KimPLB13}, and here a $\Z_{nR}$ is assumed for an explicit discussion. Let the inflaton $\Phi$ carries a negative $\Z_{nR}$ charge but let all GUT scale scalars $\psi_i$ carry positive $\Z_{nR}$ charges. The VEVs $\langle\psi_i\rangle$ are at the GUT scale. Let us assume only one $\psi_i$ for simplicity. The effective superpotential terms are obtained by assigning GUT scale VEVs to the GUT scale scalars,
\dis{
\sum_{i}\frac{\psi^{a_i}}{M_P^{a_i+\ell_i- 3}}\Phi^{\ell_i};~\textrm{with constraint}~  a_i n_\psi+\ell_i n_\Phi=2.\label{eq:Wnonren}
}
where $n_\psi>0$ and $n_\Phi<0$ are the $\Z_{nR}$ quantum numbers of $\psi$ and $\Phi$, respectively, and we have the relation $ a_i n_\psi=2-\ell_i n_\Phi$.
Since $n_\Phi<0$, we must have $a_i=(\ell_i |n_\Phi|+2)/n_\psi$,   and the most dangerous term with the minimal form of K\"ahler potential is $|\partial W/\partial\psi|^2$
\dis{
&
\sum_{i,j} \frac{a_i a_j \psi^{ a_i+a_j-2}}{M_P^{a_i+a_j+\ell_i+\ell_j-6}}\Phi^{\ell_i+\ell_j}\\
&~~=
 \sum_{i,j}a_i a_j\left(\frac{\psi }{M_P }\right)^{\frac{ (\ell_i+\ell_j)|n_\Phi|-2n_\psi+4}{n_\psi}} \left(\frac{\Phi }{M_P }\right)^{ \ell_i+\ell_j}\, M_P^4.
}
Let us take $\psi/M_P\sim 10^{-2}$ and $\Phi$ as a trans-Planckian value \cite{Lyth97}, $\Phi/M_P\sim 31\simeq 10^{3/2}$ for an illustration. Then, we estimate the magnitude of $V$ as
\dis{
 \approx \sum_{i,j}\left(10 \right)^{\log a_i+\log a_j-\frac{2(\ell_i+\ell_j)|n_\Phi|-4n_\psi +8-\frac32(\ell_i+\ell_j) n_\psi}{n_\psi}}  M_P^4.
}
Therefore, if $|n_\Phi|> \frac{3}{4} n_\psi$, we obtain successively decreasing $\Phi^\ell$ terms as $\ell$ increases for a large $\ell$ and obtain a reasonable expansion. This is in contrast to what Lyth recently commented against SUSY \cite{Lyth14}. Lyth's criteria will apply to any theory, even for models without SUSY if it needs a trans-Planckian inflaton field value, asking for a rationale of the potential cutting off the higher power inflaton terms. The trick we obtain a successively decreasing series from (\ref{eq:Wnonren}) is that as soon as the power $\ell$ of the trans-Planckian field $\Phi$ increases, the suppression from the GUT field $\psi$ increases more rapidly due the discrete symmetry constraint originating from a gauge symmetry \cite{KimPLB13}.
Of course, this method does not work outside the hilltop region.

\section{Conclusion}
Based on the recent BICEP2 report on the CMB B-mode polarization, we analysed a few implications of our recently proposed  \UDE~ hilltop inflation model. The \UDE\,hilltop inflation alone cannot describe the BICEP2 data at bull's eye, but by coupling it to a chaoton field it can successfully locate the BICEP2 point. In this case, a trans-Planckian decay constant is needed. We commented how the trans-Planckian decay constant can lead to a GUT scale energy density in the hilltop side of the potential, which is made possible from some discrete symmetries of ultra-violet completed models.

\section*{Acknowledgments}
I thank W. Buchm\"uller, L. Covi, J. Hong, K. S. Jeong, and H. P. Nilles for helpful discussions.
This work is supported in part by the National Research Foundation (NRF) grant funded by the Korean Government (MEST) (No. 2005-0093841) and  by the IBS(IBS CA1310).



\begin{thebibliography}{99}

\def\prp#1#2#3{Phys.\ Rep.\ {\bf #1} (#3) #2}
\def\rmp#1#2#3{Rev. Mod. Phys.\ {\bf #1} (#3) #2}
\def\npb#1#2#3{Nucl.\ Phys.\ {\bf B#1} (#3) #2}
\def\plb#1#2#3{Phys.\ Lett.\ {\bf B#1} (#3) #2}
\def\prd#1#2#3{Phys.\ Rev.\ {\bf D#1} (#3) #2}
\def\prl#1#2#3{Phys.\ Rev.\ Lett.\ {\bf #1} (#3) #2}
\def\jhep#1#2#3{JHEP\ {\bf #1} (#3) #2}
\def\jcap#1#2#3{JCAP\ {\bf #1} (#3) #2}
\def\zp#1#2#3{Z.\ Phys.\ {\bf #1} (#3) #2}
\def\epjc#1#2#3{Euro. Phys. J.\ {\bf C#1} (#3) #2}
\def\jpg#1#2#3{J. Phys.\ {\bf G#1} (#3) #2}
\def\ijmp#1#2#3{Int.\ J.\ Mod.\ Phys.\ {\bf #1} (#3) #2}
\def\mpl#1#2#3{Mod.\ Phys.\ Lett.\ {\bf A#1} (#3) #2}
\def\apj#1#2#3{Astrophys.\ J.\ {\bf #1} (#3) #2}
\def\nat#1#2#3{Nature\ {\bf #1} (#3) #2}
\def\sjnp#1#2#3{Sov.\ J.\ Nucl.\ Phys.\ {\bf #1} (#3) #2}
\def\apj#1#2#3{Astrophys.\ J.\ {\bf #1} (#3) #2}
\def\ijmp#1#2#3{Int.\ J.\ Mod.\ Phys.\ {\bf #1} (#3) #2}
\def\mpla#1#2#3{Mod.\ Phys.\ Lett.\ {\bf A#1} (#3) #2}
\def\nat#1#2#3{Nature\ {\bf #1} (#3) #2}
\def\npb#1#2#3{Nucl.\ Phys.\ {\bf B#1} (#3) #2}
\def\pthp#1#2#3{Prog.\ Theor.\ Phys.\ {\bf #1} (#3) #2}
\def\jkps#1#2#3{J.\ Korean Phys.\ Soc.\ {\bf #1} (#3) #2}

\def\ibid#1#2#3{{\it ibid.} {\bf #1} (#3) #2}
\def\err#1#2#3{\ {\bf #1} (#3) #2\,(E)}
\def\err#1#2#3{\ {\bf #1} (#3) #2\,(E)}

\bibitem{BICEP2I}
P.~A.~R.~Ade  \etal\,[BICEP2 Collaboration],  arXiv:1403.3985[astro-ph.CO].

\bibitem{Planck13} P. A. R. Ade \etal\,(Planck Collaboration),
 [arXiv: 1303.5082].

\bibitem{KimNilles14} J. E. Kim and H. P. Nilles, \plb{730}{53}{2014} [arXiv:1311.0012 [hep-ph]].

\bibitem{KimJK14} J. E. Kim, \jkps{64}{795}{2014} [arXiv:1311.4545[hep-ph]].

\bibitem{Buch14}  W. Buchmuller, C. Wieck, and M. W. Winkler, arXiv:1404.2275 [hep-th].

\bibitem{Busso13} R. Busso, D. Harlow, and L. Senatore, arXiv:1309.4060 [hep-th].

\bibitem{Perlmutter98} S. Perlmutter \etal (Supernova Cosmology Project),  \apj{517}{565}{1999}.

\bibitem{Riess98} A. G. Riess \etal (Supernova Search Team), \apj{116}{1009}{1998}.

\bibitem{KimPLB13} J. E. Kim, \plb{726}{450}{2013} [arXiv:1308.0322 [hep-th]].

\bibitem{Carroll98} S. M. Carroll, \prl{81}{3067}{1998} [arXiv:astro-ph/9806099].

\bibitem{InvAxionRev10}  For a recent review, see, J. E. Kim and G. Carosi,
    \rmp{82}{557}{2010} [arXiv: 0807.3125[hep-ph]].

\bibitem{Guth} A. Guth, \prd{23}{347}{1981}.

\bibitem{NewInfl} A. D. Linde, \plb{108}{389}{1982};\\
 A. Albrecht and P. J. Steinhardt, \prl{48}{1220}{1982}.

\bibitem{NaturalInfl} K. Freese, J. A. Freeman, and A. V. Orlinto, \prl{65}{3233}{1990}.

\bibitem{Peloso04} J. E. Kim, H. P. Nilles, and M. Peloso, \jcap{0501}{005}{2005} [arXiv:hep-ph/0409138].

\bibitem{Lyth97} D. Lyth, \prl{78}{1861}{1997} [hep-ph/9606387].

\bibitem{Nflation08}  S. Dimopoulos, S. Kachru, J. McGreevy, and J. G. Wacker, \jcap{0808}{003}{2008}
    [arXiv: hep-th/0507205].

\bibitem{Kim:2006hw}
  J.~E.~Kim and B.~Kyae, \npb{770}{47}{2007}   [hep-th/0608086].

\bibitem{Lebedev:2006kn}
  O.~Lebedev, H.~P.~Nilles, S.~Raby, S.~Ramos-Sanchez, M.~Ratz,
P.~K.~S.~Vaudrevange and A.~Wingerter, \plb{645}{88}{2007}
  [hep-th/0611095].

\bibitem{Kim:2006zw}
  I.~-W.~Kim, J.~E.~Kim and B.~Kyae, \plb{647}{275}{2007}
   [hep-ph/0612365].

\bibitem{Lebedev:2007hv}
  O.~Lebedev, H.~P.~Nilles, S.~Raby, S.~Ramos-Sanchez, M.~Ratz,
P.~K.~S.~Vaudrevange and A.~Wingerter, \prd{77}{046013}{2008}    [arXiv:0708.2691 [hep-th]].

\bibitem{Lebedev:2008un}
  O.~Lebedev, H.~P.~Nilles, S.~Ramos-Sanchez, M.~Ratz and P.~K.~S.~Vaudrevange, \plb{668}{331}{2008}
  [arXiv:0807.4384 [hep-th]].

\bibitem{Nilles:2008gq}
  For a review, see: H.~P.~Nilles, S.~Ramos-Sanchez, M.~Ratz and P.~K.~S.~Vaudrevange, \epjc{59}{249}{2009}  [arXiv:0806.3905 [hep-th]].

\bibitem{CW82} S. Coleman and E. Weinberg, \prd{7}{1888}{1973}.

\bibitem{WeinbergCW76} S. Weinberg, \prl{36}{294}{1976}; A. D. Linde, \plb{70}{306}{1977}.

\bibitem{Linde83} A. D. Linde, \plb{129}{177}{1983}.

\bibitem{JeongKim14} K. S. Jeong and J. E. Kim, work in progress.

\bibitem{Lyth14} D. Lyth, arXiv:1403.7323 [hep-ph].

\end{thebibliography}
\end{document}